\documentclass[doublecol]{epl2} 

\usepackage[utf8]{inputenc}
\usepackage{amsmath,amssymb}
\usepackage{graphicx}
\usepackage[colorlinks=true,citecolor=blue]{hyperref}
\usepackage{units}

\renewcommand{\vec}[1]{\mathbf{#1}}
\def\up{\uparrow}
\def\down{\downarrow}

\DeclareMathOperator{\sech}{sech}

\title{Magnon-driven quantum-dot heat engine}
\shorttitle{Magnon-driven quantum-dot heat engine} 

\author{Bj\"orn Sothmann \and Markus B\"uttiker}
\shortauthor{Bj\"orn Sothmann \etal}

\institute{                    
  Département de Physique Théorique, Université de Genève, CH-1211 Genève 4, Switzerland
}

\abstract{
We investigate a heat- to charge-current converter consisting of a single-level quantum dot coupled to two ferromagnetic metals and one ferromagnetic insulator held at different temperatures. 
We demonstrate that this nano engine can act as an optimal heat to spin-polarized charge current converter in an antiparallel geometry, while it acts as a heat to pure spin current converter in the parallel case. We discuss the maximal output power of the device and its efficiency.}

\begin{document}

\maketitle

\section{\label{sec:introduction}Introduction}
Thermoelectrics in nanostructures has generated much interest recently, notably due to the strong energy-dependence of transmissions which allows large thermoelectric efficiencies~\cite{mahan_best_1996}.
Quantum dots offer great versability and unprecedented control over their parameters, which has lead to a particular interest in their thermoelectric properties~\cite{beenakker_theory_1992,humphrey_reversible_2002,swirkowicz_thermoelectric_2009,sanchez_optimal_2011,sothmann_rectification_2012,muralidharan_performance_2012}. Experimentally, thermoelectrics has been studied both for open quantum dots~\cite{godijn_thermopower_1999,llaguno_observation_2003} and for Coulomb-blockade dots~\cite{molenkamp_sawtooth-like_1994,staring_coulomb-blockade_1993,dzurak_thermoelectric_1997}.

From a theoretical point of view, thermoelectrics in multi-terminal setups proves to be interesting as they allow for crossed flows of heat and charge currents. So far, two different types of such multi-terminal setups with quantum dots have been discussed. In the first, two quantum dots are capacitively coupled to each other thus allowing the exchange of energy but not of particles between the dots. Both dots are then coupled to hot and cold reservoirs, respectively. In the Coulomb-blockade regime, these devices have been shown to be optimal heat to charge converters that can reach Carnot efficiency~\cite{sanchez_optimal_2011}. In ref.~\cite{sothmann_rectification_2012}, the scaling with the number of quantum channels of such setups was obtained for large dots in the form of chaotic cavities.

The second class of devices consists of quantum dots coupled to two electronic reservoirs as well as to a phonon bath at a different temperature. Both, the case of a single dot~\cite{entin-wohlman_three-terminal_2010} as well as that of a two-site molecule~\cite{jiang_thermoelectric_2012} have been discussed in the framework of linear-response thermoelectrics. Additionally, the effect of time-reversal symmetry breaking has been explored for such devices where the dot is embedded in an Aharonov-Bohm ring~\cite{entin-wohlman_three-terminal_2012}.

\begin{figure}
	\onefigure[width=8cm]{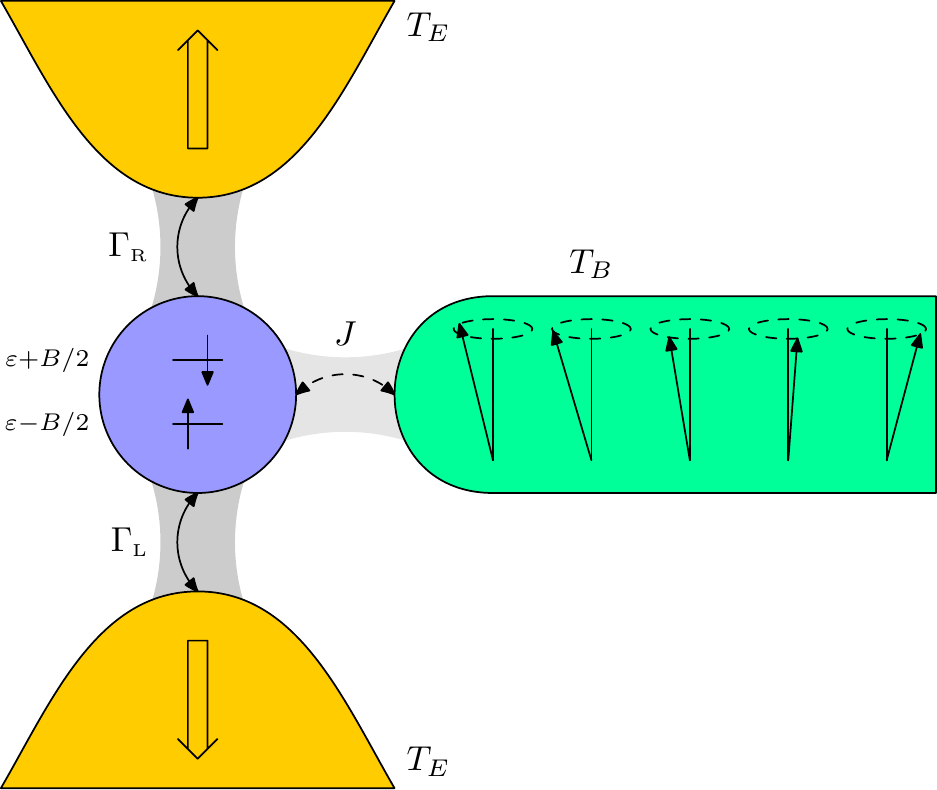}
	\caption{(Color online) Schematic of the heat-to-current converter. A quantum dot with a single spin-split level is coupled to two ferromagnetic metallic reservoirs at temperature $T_E$ and an additional ferromagnetic insulator held at a different temperature $T_B$.}
	\label{fig:model}
\end{figure}

Here, we propose a third class of setups in which a single-level quantum dot is coupled to two ferromagnetic metals at temperature $T_E$ as well as to a third ferromagnetic insulator at temperature $T_B$, cf. fig.~\ref{fig:model}. For $T_B>T_E$, the quantum dot will absorb magnons from the insulator and in turn transfer electrons between the two metallic leads.
Compared to the coupling of a phonon bath to the quantum dot, our setup offers several advantages.
First, magnons provide a more controlled system as they are restricted to the ferromagnetic insulator whereas phonons are present in any material and, hence, can lead to parasitic heat flows.
Second, one can drive our quantum dot heat engine not only by a temperature difference between magnons and electrons but also by a nonequilibrium magnon distribution that is generated by injecting magnons into the ferromagnetic insulator electrically via the spin Hall effect~\cite{kajiwara_transmission_2010}. Thus, one could operate the device at low temperatures where magnons are not thermally excited and, hence, parasitic phonons are absent as well. In addition, the large mean free path of magnons in ferromagnetic insulators of several centimeters~\cite{schneider_realization_2008} allows to separate the heat engine spatially from the point of magnon injection.
Third, our setup provides an example of a spin caloritronic~\cite{bauer_spin_2012} heat engine that allows to drive pure spin currents as well as spin-polarized charge currents by thermal gradients. Alternative spintronic heat engines have been proposed using nanowires containing domain walls\cite{bauer_nanoscale_2010,kovalev_thermomagnonic_2012}.
In ref.~\cite{sothmann_influence_2010} a two-terminal quantum-dot setup was discussed that converted a nonequilibrium magnon distribution into a directed charge current.

Finally, while the setup involving Coulomb-coupled quantum dots relies on energy-dependent transmissions and needs them to be especially tailored to reach the tight-coupling limit where Carnot efficiency can be achieved~\cite{sanchez_optimal_2011}, our proposal works with energy-independent tunnel couplings and the tight-coupling limit can be achieved for completely polarized metallic electrodes.

\section{\label{sec:model}Model}
We consider a single-level quantum dot coupled to two ferromagnetic metallic reservoirs at temperature $T_E$ and one ferromagnetic insulator at temperature $T_B$. The Hamiltonian of the system consists of five parts, 
\begin{equation}
	H=\sum_r H_r+H_\text{B}+H_\text{dot}+\sum_rH_{\text{tun},r}+H_\text{tun,B}.
\end{equation}
The first part describes the two ferromagnetic metals $r=\text{L,R}$,
\begin{equation}
	H_r=\sum_{\vec k\sigma}\varepsilon_{r\vec k\sigma}a_{r \vec k\sigma}^\dagger a_{r \vec k\sigma}.
\end{equation}
Here, $a_{r\vec k\sigma}^\dagger$ creates an electron with momentum $\vec k$ and spin $\sigma$ in lead $r$. 
Ferromagnetism is incorporated in the sense of a simple Stoner model using spin-dependent densities of states, $\rho_{r\sigma}(\omega)=\sum_{\vec k}\delta(\omega-\varepsilon_{r\vec k\sigma})$. In the following, we assume these to be energy independent, $\rho_{r\sigma}(\omega)=\rho_{r\sigma}$. The spin-dependence of $\rho_{r\sigma}$ is characterized by the polarization $p_r=(\rho_{r\up}-\rho_{r\down})/(\rho_{r\up}+\rho_{r\down})$ which varies between $p_r=-1$ and $p_r=1$. The two extreme cases correspond to half-metallic ferromagnets with majority spins only while $p_r=0$ describes an unpolarized electrode.

The ferromagnetic insulator is described in terms of a Heisenberg chain
\begin{equation}\label{eq:HamiFI}
	H_\text{FI}=-\frac{\tilde J}{2}\sum_{\langle i,j\rangle}\vec S_i\cdot \vec S_j+\sum_i B_z S_{iz},
\end{equation}
where $\langle i,j\rangle$ indicates summation over nearest neighbors.
We assume the magnetization of the ferromagnetic ground state to point along the $+z$ direction and the weak external field $B_z$ to be either parallel or antiparallel to the magnetization.
We perform a Holstein-Primakoff transformation and replace the spin by bosonic operators,
\begin{align}
	S_{i+}&=\sqrt{2S-b_i^\dagger b_i}b_i,\\
	S_{i-}&=b_i^\dagger\sqrt{2S-b_i^\dagger b_i},\\
	S_{iz}&=S-b_i^\dagger b_i.
\end{align}
Assuming the length $S$ of the spin to be much larger than the average boson number $\langle b_i^\dagger b_i\rangle$, which is a reasonable approximation at low temperatures, we can expand the square roots in $1/S$ and retain only the leading order terms $S_{i+}=\sqrt{2S}b_i$, $S_{i-}=\sqrt{2S}b_i^\dagger$. Inserting into eq.~\eqref{eq:HamiFI} and performing a Fourier transformation, we obtain the Hamiltonian describing the spin wave degrees of freedom,
\begin{equation}
	H_\text{B}=\sum_{\vec q}\omega_{\vec q}b_{\vec q}^\dagger b_{\vec q}
\end{equation}
with $\omega_q=2\tilde JS(1-\cos \vec q\cdot \vec a)+B_z$ where $\vec a$ denotes a lattice vector of the ferromagnetic insulator.

The quantum dot is described by
\begin{equation}
	H_\text{dot}=\sum_\sigma \varepsilon_\sigma c_\sigma^\dagger c_\sigma+Uc_\up^\dagger c_\up c_\down^\dagger c_\down.
\end{equation}
Here, $\varepsilon_\sigma=\varepsilon\mp B/2$ denotes the energy of the spin-split level. The externally applied magnetic field $B$ is measured in units of $g\mu_\text{B}$ where $g$ is the $g$ factor and $\mu_\text{B}$ the Bohr magneton. The Zeeman splitting determines the energy of the magnons that can be absorbed/emitted by the quantum dot. In the following, we assume $B>0$ as only in this case, both energy and angular momentum conservation can be obeyed when absorbing/emitting a magnon. The Coulomb energy $U$ is the energy required to occupy the quantum dot with two electrons at the same time. In the following discussion, we assume it to be infinitely large, $U\to\infty$, such that double occupancy of the dot is forbidden.

The tunneling between the quantum dot and the ferromagnetic metals is given by
\begin{equation}
	H_{\text{tun},r}=\sum_{\vec k\sigma}t_ra_{r\vec k\sigma}^\dagger c_\sigma+\text{H.c.}
\end{equation}
The tunnel matrix elements $t_r$ are related to the spin-dependent tunnel coupling strengths via $\Gamma_{r\sigma}=2\pi|t_r|^2\rho_{r\sigma}/2$. For later convenience, we also introduce the total tunnel coupling strength $\Gamma_r=\Gamma_{r\up}+\Gamma_{r\down}$.

Finally, the coupling between the quantum dot and the ferromagnetic insulator is given by an exchange interaction which in terms of the magnonic operators takes the form
\begin{equation}
	H_{\text{tun},\text{B}}=\sum_{\vec q} j_{\vec q} b_{\vec q}^\dagger c_\up^\dagger c_\down+\text{H.c.}
\end{equation}
Here, we neglected terms of the form $(S-b_{\vec q}^\dagger b_{\vec q})c_\sigma^\dagger c_\sigma$ arising from the $z$ component of the scalar product between the dot and ferromagnetic insulator spin as for $\langle b_{\vec q}^\dagger b_{\vec q}\rangle\ll S$ they can be absorbed into the dot Hamiltonian. We, furthermore, define the spectral weight $J(\omega)=2\pi\sum_{\vec q} |j_{\vec q}|^2\delta(\omega-\omega_{\vec q})$. In contrast to the tunnel coupling strengths, $J(\omega)$ will in general have a nontrivial energy-dependence that depends, e.g., on the dimension of the spin chain~\cite{strelcyk_magnetischer_2005}. However, as for the following discussion this energy-dependence of $J(\omega)$ is irrelevant, we will assume it to be constant and write $J(\omega)=J$.

\section{\label{sec:technique}Technique}
In order to evaluate the transport properties of the quantum dot system, we use a real-time diagrammatic approach~\cite{konig_zero-bias_1996,konig_resonant_1996} in its extensions to systems with ferromagnetic~\cite{braun_theory_2004} and magnonic~\cite{strelcyk_magnon_2005} reservoirs. The idea of this approach is to split the system into the strongly interacting quantum dot region with only few degrees of freedom and the noninteracting reservoirs with many degrees of freedom. The reservoirs are integrated out using Wick's theorem. The remaining quantum dot system is then described in terms of its reduced density matrix $\rho$. In the stationary state, $\rho$ obeys a master equation of the form $0=W\rho$. The generalized transition rates $W$ can be calculated perturbatively in the tunnel coupling to the reservoirs. 

As we deal with a setup with collinear magnetizations only, the reduced density matrix is diagonal and can be written as $\rho=(P_0,P_\up,P_\down)$ with the probabilities to find the dot being empty ($P_0$) or occupied with spin up ($P_\up$) or spin down ($P_\down$). To lowest order in the coupling to the reservoirs, the master equation then takes the form
\begin{widetext}
\begin{equation}\label{eq:MasterEquation}
	0=\sum_r\left(
	\begin{array}{ccc}
		-\Gamma_{r\up}f^+_r(\varepsilon-\frac{B}{2})-\Gamma_{r\down}f^+_r(\varepsilon+\frac{B}{2}) & \Gamma_{r\up}f^-_r(\varepsilon-\frac{B}{2}) & \Gamma_{r\down}f^-_r(\varepsilon+\frac{B}{2}) \\
		\Gamma_{r\up}f^+_r(\varepsilon-\frac{B}{2}) & -\Gamma_{r\up}f^-_r(\varepsilon-\frac{B}{2})-\frac{J}{2}n^+(B) & \frac{J}{2}n^-(B) \\
		\Gamma_{r\down}f^+_r(\varepsilon+\frac{B}{2}) & \frac{J}{2}n^+(B) & -\Gamma_{r\down}f^-_r(\varepsilon+\frac{B}{2})-\frac{J}{2}n^-(B)
	\end{array}
	\right)
	\left(
	\begin{array}{c}
		P_0 \\
		P_\up \\
		P_\down
	\end{array}
	\right)
\end{equation}
\end{widetext}
\begin{floatequation}
	\mbox{\textit{see eq.~\eqref{eq:MasterEquation}.}}
\end{floatequation}
Here, $f_r^+(x)=[\exp((x-\mu_r)/T_E)+1]^{-1}$ denotes the Fermi function of lead $r$ with chemical potential $\mu_r$ and $f_r^-(x)=1-f_r^+(x)$. Furthermore, $n^+(x)=[\exp(x/T_B)-1]^{-1}$ denotes the Bose function and $n^-(x)=1+n^+(x)$. 

In order to calculate the (particle) currents and current correlations, we make use of the framework of full-counting statistics adopted for systems that can be described by master equations~\cite{bagrets_full_2003,braggio_full_2006,flindt_counting_2008}. To this end, we introduce counting fields $\boldsymbol \chi=(\chi_{\up\text{L}},\chi_{\down\text{L}},\chi_{\up\text{R}},\chi_{\down\text{R}},\chi_{B})$ for electrons with spin up (down) leaving the left (right) reservoir and magnons leaving the ferromagnetic insulator, respectively. We multiply each rate in eq.~\eqref{eq:MasterEquation} by a factor $e^{iN_{js}\chi_{js}}$ where $N_{js}$ is the number of particles of type $j$ that left reservoir $s=\text{L},\text{R},\text{B}$ in the corresponding transition to obtain the new matrix $W^\chi$ (For simplicity, we omit the index $j$ for magnons). The cumulant generating function $\mathfrak S(\boldsymbol\chi)$ is obtained as the smallest eigenvalue of $W^\chi$.
We checked numerically that it exhibits the symmetry $\mathfrak S(\boldsymbol\chi)=\mathfrak S(-\boldsymbol\chi-\boldsymbol A_V-\boldsymbol A_T)$, where $\boldsymbol A_V=i(eV_\text{L}/T_E,eV_\text{L}/T_E,eV_\text{R}/T_E,eV_\text{R}/T_E,0)$ and $\boldsymbol A_T=-i((\varepsilon-B/2)/T_E,(\varepsilon+B/2)/T_E,(\varepsilon-B/2)/T_E,(\varepsilon+B/2)/T_E,B/T_B)$, from which fluctuation relations between higher-order nonlinear cumulants can be derived~\cite{andrieux_fluctuation_2007}. Interestingly, the above relation holds without changing the direction of the reservoir magnetizations and magnetic fields as such a reversal gives rise to a system where all charge (spin) currents flow in the same (opposite) direction as for the unreversed system.
From the cumulant generating function, we obtain the currents $I_{jr}$ and the zero-frequency current correlations $S_{I_{jr}I_{j'r'}}=\int\,\upd t[\langle \hat I_{jr}(t)\hat I_{j'r'}(0)\rangle-\langle \hat I_{jr}(t)\rangle\langle \hat I_{j'r'}(0)\rangle]$ by taking derivatives with respect to the counting fields $\chi_{jr}$,
\begin{equation}
	I_{jr}=(-i)\left.\frac{\partial \mathfrak S(\boldsymbol \chi)}{\partial \chi_{jr}}\right|_{\boldsymbol\chi=0},
\end{equation}
\begin{equation}\label{eq:correlations}
	S_{I_{jr}I_{j'r'}}=(-i)^2\left.\frac{\partial \mathfrak S(\boldsymbol\chi)}{\partial \chi_{jr}\chi_{j'r'}}\right|_{\boldsymbol\chi=0}.
\end{equation}
From the spin-resolved electron currents we obtain the charge and spin currents as $I_{r}=\sum_\sigma I_{r\sigma}$ and $I_{r}^S=I_{r\up}-I_{r\down}$, respectively.

In order to calculate heat currents, we multiply the transition rates in the master equation~\eqref{eq:MasterEquation} by factors $e^{iE_{jr}\xi_{jr}}$ where $E_{jr}$ is the energy transferred in the corresponding tunneling event, measured relative to the chemical potential of lead $r$. As above, we obtain the cumulant generating function $\mathfrak S(\boldsymbol\xi)$ and thereby the heat current $J_{jr}$ as
\begin{equation}\label{eq:heatcurrent}
	J_{jr}=(-i)\left.\frac{\partial \mathfrak S(\boldsymbol \xi)}{\partial \xi_{jr}}\right|_{\boldsymbol\xi=0}.
\end{equation}

\section{Results}
In the following, we consider a symmetric system, i.e., we assume the coupling to the two ferromagnetic leads to be of equal strength, $\Gamma_\text{L}=\Gamma_\text{R}\equiv\Gamma$. We furthermore assume equal polarizations $p_\text{L}=|p_\text{R}|\equiv p>0$ and a symmetrically applied bias voltage, $V_\text{L}=-V_\text{R}\equiv V/2$.

\begin{figure}
	\includegraphics[width=\columnwidth]{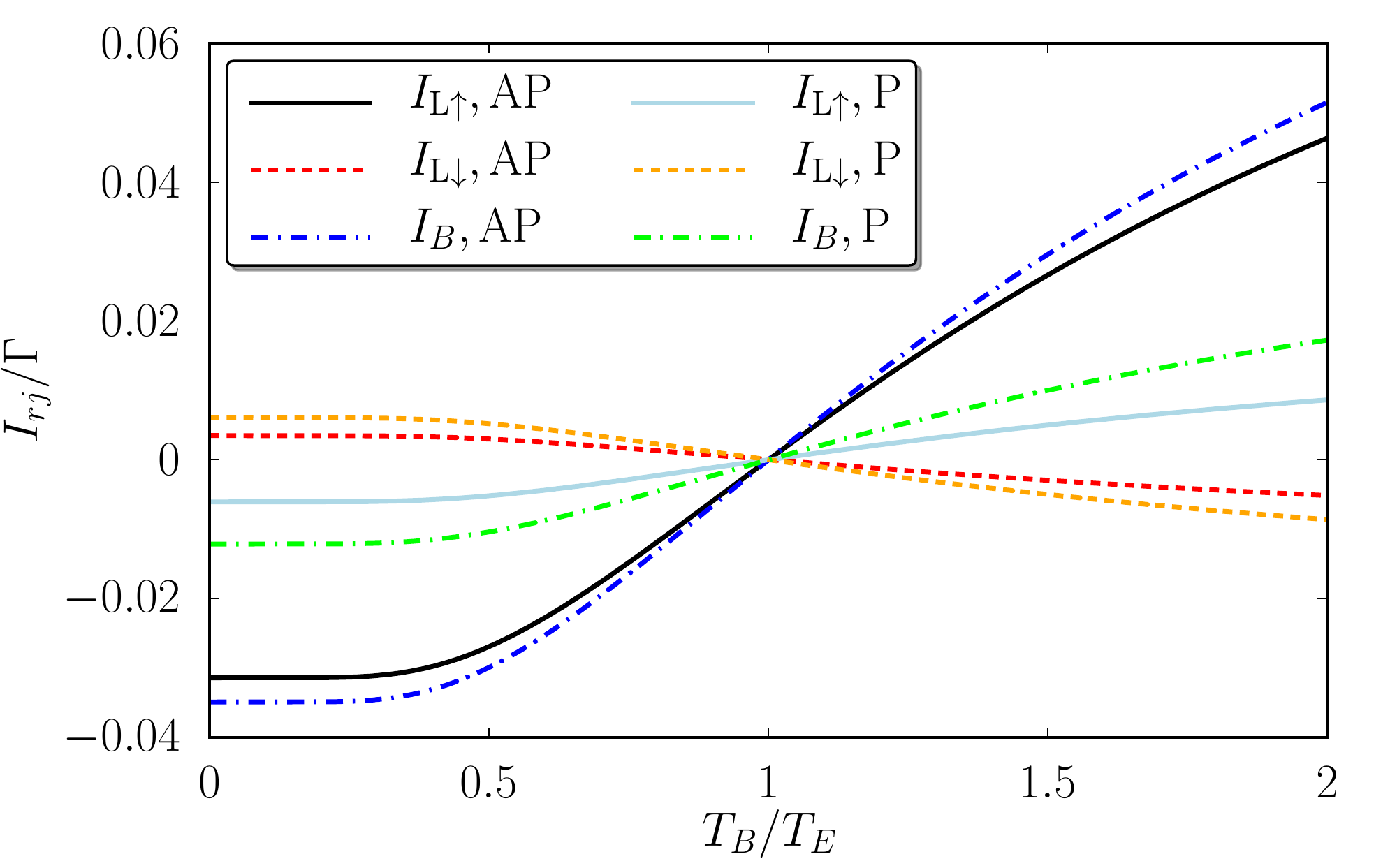}
	\caption{\label{fig:Current_TB}(Color online) Spin-resolved electron and magnon currents as a function of the magnon temperature. Parameters are $J=\Gamma$, $B=2T_E$, $\varepsilon=0$ and $p=0.8$.}
\end{figure}

We start by analyzing the spin-resolved electron and magnon currents in the absence of a bias voltage. We first discuss the basic sequence of transport processes. If the dot is initially empty, a spin-up electron enters from one of the two ferromagnetic electrodes. Afterwards, a magnon is absorbed, flipping the spin of the electron and increasing its energy by $B$. In a next step, the spin-down electron can tunnel out of the dot into the ferromagnetic electrodes. Similarly, the reversed process of injecting a high-energy spin-down electron, creating a magnon and ejecting a spin-up electron is also possible. In thermal equilibrium, $T_B=T_E$, both contributions precisely cancel and there are no average currents flowing, cf. Fig~\ref{fig:Current_TB}. If $T_B>T_E$, the absorption of magnons is more likely than their emission. In consequence, we have a magnon current flowing out of the ferromagnetic insulator, a spin-up electron current entering the dot and a spin-down current leaving the dot. If $T_B<T_E$, the situation is reversed and a magnon current flows into the ferromagnetic insulator while a spin-down electron current enters the dot and a spin-up electron current leaves it, cf. fig.~\ref{fig:Current_TB}.
We find this picture confirmed by the current cross correlations: They are positive for magnon and spin up currents while they are negative for magnon and spin down currents.

In the case of parallel magnetizations, the ratio between processes where electrons tunnel in from the left or right lead is the same as for processes where they tunnel out to the left or right lead. Hence, there is no net charge current flowing. Nevertheless, we find a pure spin current flowing out of the quantum dot into both leads. As the polarization is increased, this current is reduced because the tunneling out rate for spin down electrons becomes suppressed with $1-p$. In the antiparallel case, on which we will focus from now on, spin-up electrons enter preferrably from the left while spin-down electrons leave preferrably to the right. Therefore, we now find a finite, spin-polarized charge current through the quantum dot. Analytically, we obtain the magnon current
\begin{widetext}
\begin{equation}\label{eq:current}
	I_B=J\Gamma\frac{f^+(\varepsilon+\frac{B}{2})\left[f^-(\varepsilon-\frac{B}{2})+n^+(B)\right]-f^+(\varepsilon-\frac{B}{2})n^+(B)}{f^+(\varepsilon+\frac{B}{2})\left[\Gamma f^+(\varepsilon-\frac{B}{2})-Jn^-(B)\right]-\left[\Gamma+J\left\{1+\left[2+f^+(\varepsilon-\frac{B}{2})\right]\right\}n^+(B)\right]}
\end{equation}
\end{widetext}
\begin{floatequation}
	\mbox{\textit{see eq.~\eqref{eq:current},}}
\end{floatequation}
while the spin-resolved electron currents are given by $I_{\text{L}\up}=\frac{1+p}{2}I_B$, $I_{\text{L}\down}=-\frac{1-p}{2}I_B$, $I_{\text{R}\up}=-I_{\text{L}\down}$ and $I_{\text{R}\down}=-I_{\text{L}\up}$. The spin-up current through the left barrier increases linearly with the polarization $p$. The spin-down current, which has opposite sign, does so as well and vanishes for $p=1$. Hence, the larger the polarization, the larger the charge current flowing through the system. For $p=1$ we are in the tight-coupling limit where each absorbed magnon gives rise to the transfer of one electron from the left to the right electrode. For realistic parameters of $\Gamma\sim J\sim\unit[1]{GHz}$, we expect currents of the order of a few pA.

We next turn to the discussion of the Onsager relations~\cite{onsager_reciprocal_1931,casimir_onsagers_1945,callen_application_1948} for our setup. Expanding the charge current through the left tunnel barrier and the heat current carried by the magnons to lowest order in the applied bias voltage $\Delta V$ and the heat gradient $\Delta T=T_B-T_E$,
\begin{align}
	I_\text{L}&=G\Delta V+L\Delta T,\\
	J_\text{B}&=M\Delta V+N\Delta T,
\end{align}
we find that the Onsager relation $M=T_E L$ is fulfilled with 
\begin{equation}
	L=\frac{pB\Gamma J\sech\frac{B}{4T_E}}{4T_E^2\left(2J\cosh\frac{B}{4T_E}+J\cosh\frac{3B}{4T_E}+\Gamma\sinh\frac{3B}{4T_E}\right)}.
\end{equation}
Here, we set $\varepsilon=0$ for simplicity. We note, however, the Onsager relation is satisfied for any value of $\varepsilon$.
We remark that for the Onsager relation to hold we have to switch both the magnetizations and magnetic fields (which for our system is equivalent to not switching them at all, cf. the discussion above). Reverting only the direction of the magnetic field ($B\to-B$) leads to an apparent violation of the Onsager relation as the coefficients $L$ and $M$ are neither even nor odd functions of $B$.

\begin{figure}
	\includegraphics[width=\columnwidth]{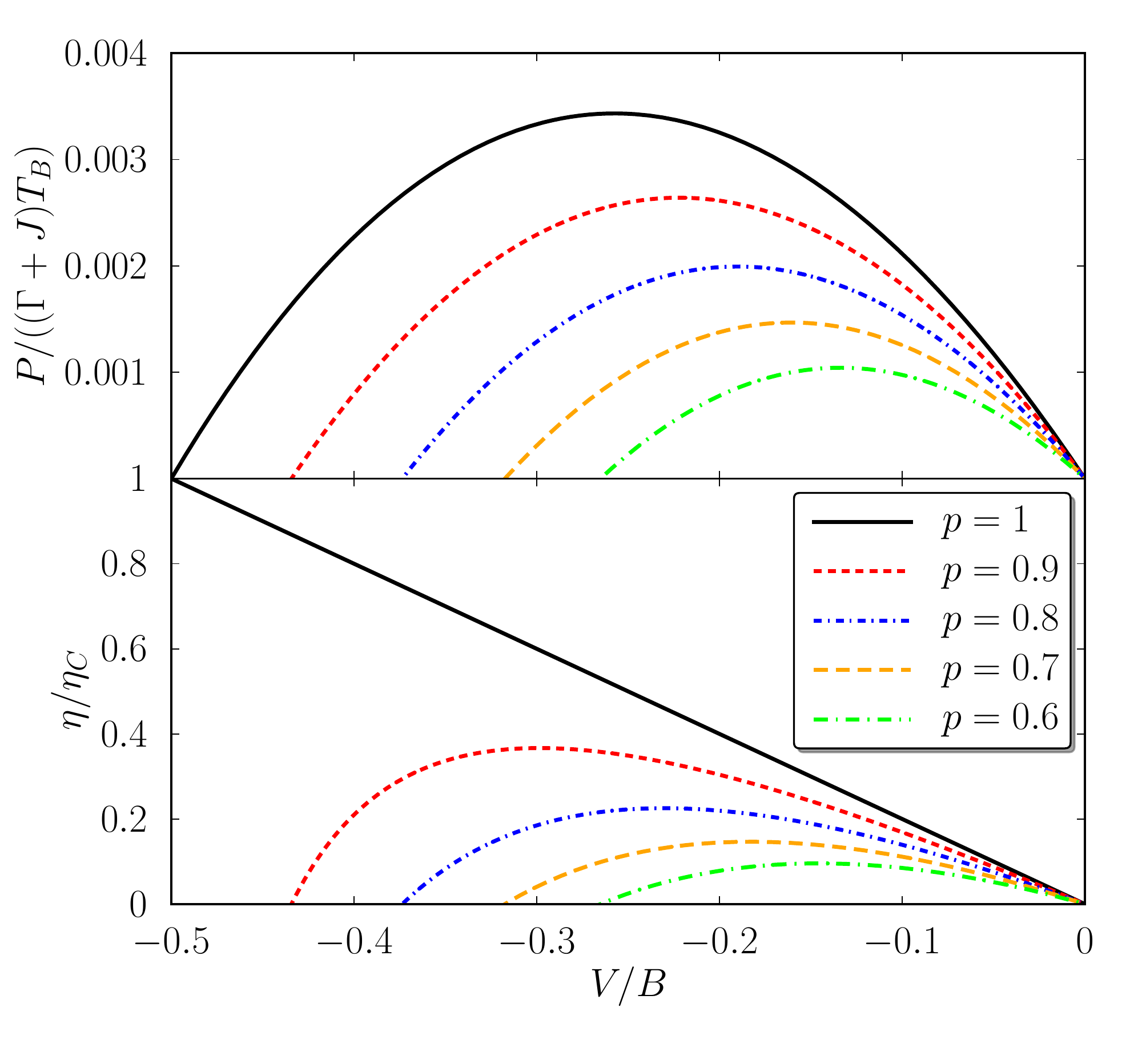}
	\caption{\label{fig:PowerEfficiency}(Color online) Power and efficiency as a function of the applied voltage for different polarizations in the antiparallel setup. Parameters are $J=\Gamma$, $B=2T_E$, $\varepsilon=0$, $T_B=2T_E$.}
\end{figure}
We now turn to the power that can be extracted from the device in the antiparallel geometry. In order to do useful work, we need to attach an external load to our system. To this end, we apply a finite bias voltage against which the thermoelectric current can do work. The output power is then simply given by $P=I_\text{L} V$.

In fig.~\ref{fig:PowerEfficiency}, we show the output power as a function of the external bias voltage. For $V=0$ the output power vanishes as there is no load to work against. For increasing bias, the power increases, reaches a maximum and then goes to zero again at the stopping potential $V_\text{stop}$ at which $I_\text{L}=0$, i.e., the thermoelectric current is exactly compensated by the voltage-induced current at this point. We, furthermore, see that as the polarization is increased, the output power also increases. This is a direct consequence of the magnon current being converted more efficiently into a charge current for large polarizations which culminates in the conversion of one magnon into one transferred electron for $p=1$. We estimate output powers of $\sim\unit[1]{fW}$ for $\Gamma\sim\unit[1]{Ghz}$ and $B\sim\unit[0.1]{meV}$.

We next turn to the efficiency of heat to work conversion which is defined as the ratio between the output power and the input heat. Hence, for $T_B>T_E$ where heat flows from the insulator into the electronic system, we have $\eta=P/J_B$ while for $T_E>T_B$ where heat flows from the electronic system into the insulator, we have $\eta=P/\sum_{r\sigma}J_{\sigma r}$.

The efficiency as a function of the applied bias voltage is shown in the lower panel of fig.~\ref{fig:PowerEfficiency}. For $p<1$, we see that $\eta=0$ at zero bias and the stopping potential as in these cases there is no voltage and no current, respectively, and, hence, no output power. In between, the efficiency reaches a maximum that increases with the polarization $p$. We note the maximal efficiency in general occurs at a different bias voltage than the maximal power.

The efficiency behaves rather differently for $p=1$. Here, it again vanishes at $V=0$ but then increases linearly with the applied bias voltage, reaching Carnot efficiency $\eta_\text{C}$ at the stopping potential. This is a consequence of the tight-coupling limit reached for $p=1$ where heat and charge currents are directly proportional to each other~\cite{esposito_universality_2009}. However, it is important to realize that for a heat engine at Carnot efficiency, no work can be extracted. In fact, at the stopping potential we have vanishing output power in combination with vanishing heat currents.

\begin{figure}
	\includegraphics[width=\columnwidth]{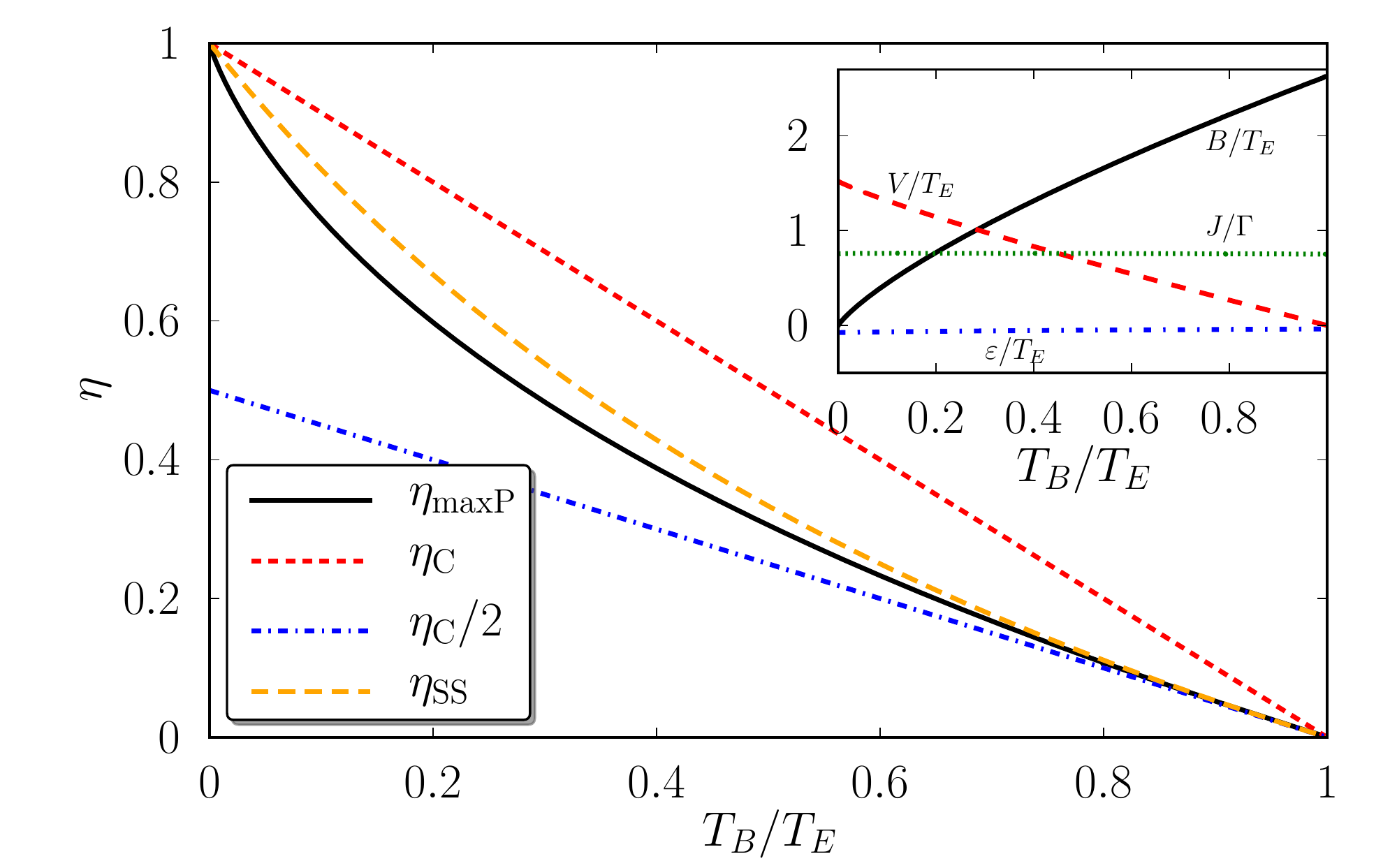}
	\includegraphics[width=\columnwidth]{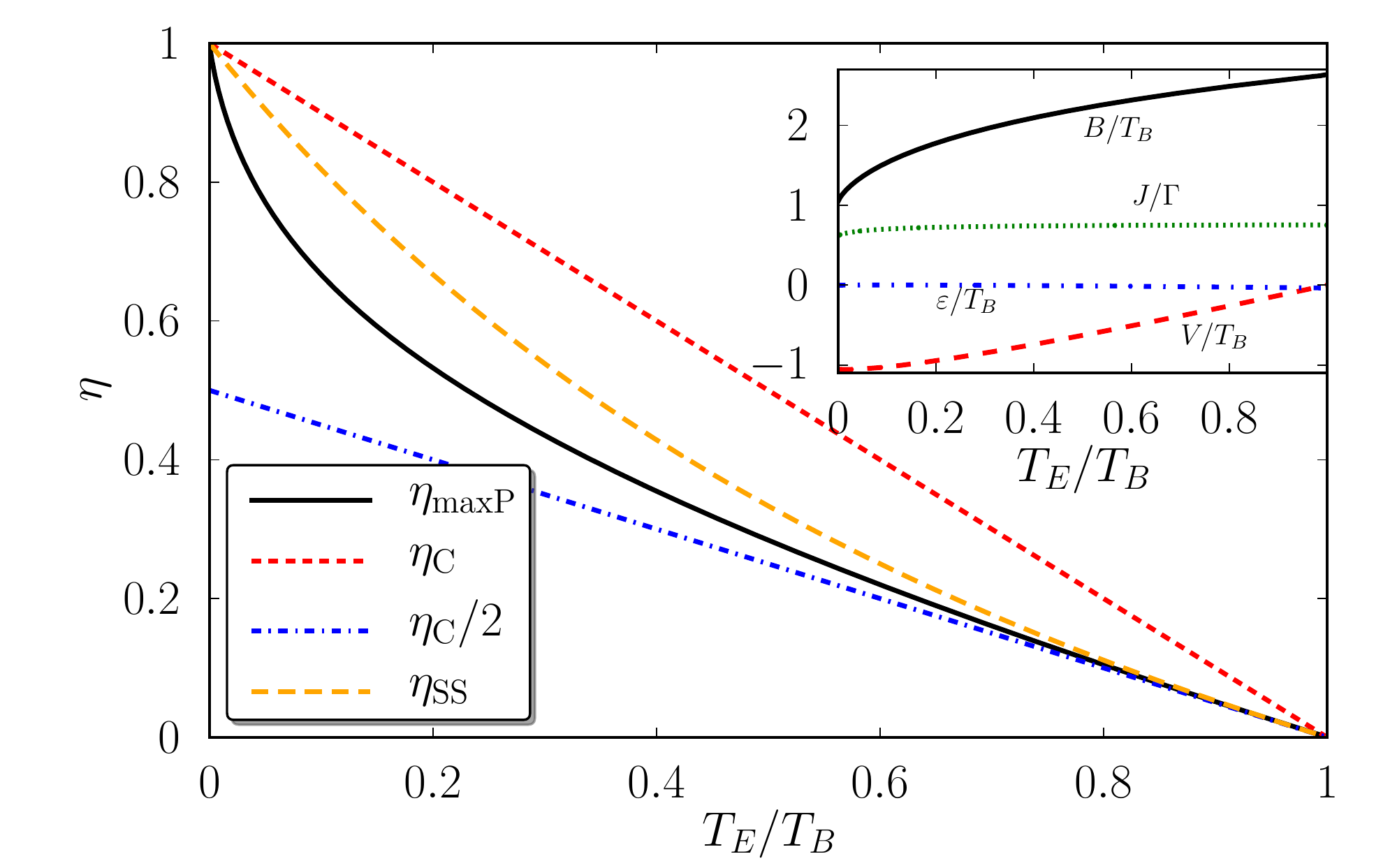}
	\caption{\label{fig:MaxEff}(Color online) Efficiency at maximum power as a function of $T_B$ (upper panel) and $T_E$ (lower panel) for  $p=1$ in the antiparallel geometry. Insets show optimized values of $\varepsilon$, $B$ and $V$. The optimization was carried out for the dimensionless quantity $P/[(J+\Gamma)T_{E/B}]$ for the upper and lower panel, respectively.}
\end{figure}

The more relevant quantity to judge the efficiency of our device is the efficiency at maximum power $\eta_\text{maxP}$ which is shown in fig.~\ref{fig:MaxEff} together with the parameters $\varepsilon$, $B$, $V$ and $J$ that maximize the output power for both directions of operation. Close to thermal equilibrium, $T_B=T_E$, we find that $\eta_\text{maxP}=\eta_\text{C}/2$ in agreement with general thermodynamic arguments~\cite{van_den_broeck_thermodynamic_2005}. We remark that this relation usually does not hold for systems whose thermopower is not symmetric under time reversal~\cite{benenti_thermodynamic_2011}. As we deviate from $T_B=T_E$, we note that $\eta_\text{maxP}$ becomes larger than $\eta_\text{C}/2$ and reaches Carnot efficiency when the temperature of the cold bath vanishes. A similar behavior was found in Ref.~\cite{esposito_thermoelectric_2009} where the nonlinear thermoelectric efficiency of a quantum dot was discusssed for the first time. Interestingly, $\eta_\text{maxP}$ is larger when the electron temperature is larger than the magnon temperature. The efficiency at maximum power satisfies the bounds $\eta_\text{C}/2<\eta_\text{maxP}<\eta_\text{C}/(2-\eta_\text{C})$. This is in agreement with the findings of  Schmiedl and Seifert\cite{schmiedl_efficiency_2008} who gave the upper bound $\eta_\text{SS}=\eta_\text{C}/(2-\eta_\text{C})$ for the efficiency at maximum power.

\section{Conclusions}
We demonstrated that a single-level quantum dot coupled to two ferromagnetic metals and a ferromagnetic insulator can act as a heat to current converter. Depending on the magnetic configuration of the metalic reservoirs, either a pure spin current or a spin-polarized charge current is generated. We showed the validity of Onsager relations for heat and charge currents. We analyzed the maximal output power and demonstrated that in the tight-coupling limit the device can reach Carnot efficiency. Finally, we discussed the efficiency at maximum power of the magnon-driven quantum dot heat engine and found that it satisfies the bounds $\eta_\text{C}/2<\eta_\text{maxP}<\eta_\text{SS}$.

\acknowledgments
We thank A. N. Jordan, J. König and R. Sánchez for valuable comments on the manuscript. We acknowledge financial support from the Swiss NSF and the project NANOPOWER (FP7/2007-2013) under Grant No. 256959.


\end{document}